\documentclass[numbers,square,sort&compress,final,twocolumn]{elsarticle}
\journal{Nucl. Instrum. Methods Phys. Res. A}
\graphicspath{{figures/}}

\usepackage{amsmath,amssymb}
\usepackage{mathrsfs}
\usepackage[colorlinks=true]{hyperref}
\usepackage{booktabs}

\usepackage{siunitx}
\usepackage{subfigure}
\usepackage{soul}

\newcommand{\DM}{\mbox{DarkMESA}}
\newcommand{\pe}{\mbox{p.e.}} % detected photoelectrons
\newcommand{\ph}{\mbox{ph.}} % detected photons

\begin{document}
%========================================================

%--------------------------------------------------------
\begin{frontmatter}
%--------------------------------------------------------

  \title{ Electron beam studies of light collection\\
    in a scintillating counter with embedded fibers}

  \date{today}

  \author[KPH]{M.~Lau{\ss}\fnref{Master}} % MAGIX Mainz group & thesis
  \author[KPH,HIM,PRISMA]{P.~Achenbach\corref{corr}} % MAGIX Mainz group & corresponding author
  \ead{achenbach@uni-mainz.de} 
  \author[KPH]{S.~Aulenbacher} % MAGIX Mainz group
  \author[HISKP]{M.~Ball} % MAGIX Bonn
  \author[KPH]{I.~Beltschikow} % KPH Contributor
%  \author[StonyBrook,Riken]{J.~C.~Bernauer} % MAGIX Stony Brook group
  \author[KPH]{M.~Biroth} % MAGIX Mainz group
  \author[Munster]{P.~Brand} % MAGIX Münxster group
  \author[KPH]{S.~Caiazza} % MAGIX Mainz group
  \author[KPH,HIM]{M.~Christmann} % MAGIX Mainz group
%  \author[StonyBrook]{E.~Cline} % MAGIX Stony Brook group
  \author[KPH]{O.~Corell} % KPH Contributor
  \author[KPH,HIM,PRISMA]{A.~Denig} % MAGIX Mainz group
  \author[KPH,PRISMA]{L.~Doria} % MAGIX Mainz group
  \author[KPH]{P.~Drexler} % KPH Contributor
%  \author[MIT]{I.~Fri\v{s}\v{c}i\'c} % MAGIX MIT group
  \author[KPH]{J.~Geimer} % MAGIX Mainz group
  \author[KPH]{P.~G\"ulker} % MAGIX Mainz group
%  \author[Munster]{A.~Khoukaz} % MAGIX Münxster group
%  \author[Hampton]{M.~Kohl} % MAGIX Hampton group
  \author[JSI]{T.~Kolar} % MAGIX Ljubljana group
  \author[KPH]{W.~Lauth} % KPH Contributor
  \author[KPH]{M.~Littich} % MAGIX Mainz group
  \author[PI]{M.~Lupberger} % MAGIX Bonn
  \author[KPH]{S.~Lunkenheimer} % MAGIX Mainz group
  \author[KPH]{D.~Markus} % MAGIX Mainz group
  \author[HIM]{M.~Mauch} % MAGIX Mainz group
  \author[KPH,PRISMA]{H.~Merkel} % MAGIX Mainz group
  \author[JSI,Ljubljana]{M.~Mihovilovi\v{c}} % MAGIX Ljubljana group
%  \author[MIT]{R.~G.~Milner} % MAGIX MIT group
  \author[KPH]{J.~M\"uller} % MAGIX Mainz group
  \author[KPH]{B.~S.~Schlimme} % MAGIX Mainz group
  \author[KPH,PRISMA]{C.~Sfienti} % MAGIX Mainz group
  \author[JSI,Ljubljana]{S.~\v{S}irca} % MAGIX Ljubljana group
%  \author[Katowice]{E.~Stephan} % MAGIX Katowice group
  \author[KPH]{S.~Stengel} % MAGIX Mainz group
  \author[KPH]{C.~Szyszka} % MAGIX Mainz group
  \author[Munster]{S.~Vestrick} % MAGIX Münxster group
%  \author[MIT]{Y.~Wang} % MAGIX MIT group
%  \author[Katowice]{A. Wilczek} % MAGIX Katowice group
  \author{\\for the MAGIX Collaboration}

  \fntext[Master]{Part of master thesis.}
%  \fntext[PhD]{Part of doctoral thesis.}
  \cortext[corr]{Corresponding author at: Institut f\"ur Kernphysik,
    J.-J.-Becherweg 45, Johannes Gutenberg-Universit\"at, D-55099
    Mainz, Germany. Tel.: +49 6131 39 25777; Fax: +49 6131 39 22964.}

  \address[KPH]{Institut f\"ur Kernphysik, Johannes
    Gutenberg-Universit\"at,  J.-J.-Becherweg 45, D-55099 Mainz, Germany}

  \address[HIM]{Helmholtz Institute Mainz, GSI Helmholtzzentrum für
    Schwerionenforschung, Darmstadt, Johannes Gutenberg-Universit\"at,
    D-55099 Mainz, Germany}

  \address[PRISMA]{PRISMA$^+$ Cluster of Excellence, Johannes
    Gutenberg-Universit\"at, Staudingerweg 9, D-55128 Mainz, Germany}

  \address[HISKP]{Helmholtz-Institut f\"ur Strahlen- und Kernphysik,
    Rheinische Friedrich-Wilhelms-Universit\"at Bonn, Nussallee
    14--16, D-53115 Bonn, Germany}

  % \address[StonyBrook]{Center for Frontiers in Nuclear Science,
  % Stony Brook University, Stony Brook, NY, USA}

  % \address[Riken]{Riken BNL Research Center, Upton, NY, USA}

  \address[Munster]{Institut f\"ur Kernphysik, Westf\"alische
    Wilhelms-Universit\"at M\"unster, Wilhelm-Klemm-Str. 9, D-48149
    M\"unster, Germany}

  % \address[MIT]{Laboratory for Nuclear Science, Massachusetts
  % Institute of Technology, Cambridge, Massachusetts 02139, USA}

  % \address[Hampton]{Department of Physics, Hampton University,
  % Hampton, Virginia 23668, USA}

  \address[JSI]{Jo\v{z}ef Stefan Institute, SI-1000 Ljubljana, Slovenia}

  \address[Ljubljana]{Faculty of Mathematics and Physics, University
    of Ljubljana, SI-1000 Ljubljana, Slovenia}

  \address[PI]{Physikalisches Institut, Rheinische
    Friedrich-Wilhelms-Universit\"at, Nussallee 12, D-53115 Bonn,
    Germany}

  % \address[Katowice]{Institute of Physics, University of Silesia,
  % 40007, Katowice, Poland}

  \hypersetup{pdfauthor={MAGIX Collaboration}}

  \begin{abstract}
    The light collection of several fiber configurations embedded in a
    box-shaped plastic scintillating counter was studied by scanning
    with minimum ionizing electrons. The light was read out by silicon
    photomultipliers at both ends. The light yield produced by the
    \SI[number-unit-product=-]{855}{\MeV} beam of the Mainz Microtron
    showed a strong dependence on the transverse distance from the
    beam position to the fibers. The observations were modeled by
    attributing the collection of indirect light inside of the counter
    and of direct light reaching a fiber to the total light yield. The
    light collection with fibers was compared to that of a
    scintillating counter without fibers. These studies were carried
    out within the development of plastic scintillating detectors as
    an active veto system for the \DM\ electron beam-dump experiment
    that will search for light dark matter particles in the \si{\MeV}
    mass range.
  \end{abstract}


%--------------------------------------------------------
\end{frontmatter}
%--------------------------------------------------------

%\linenumbers

% --------------------------------------------------------
\section{Introduction}
%--------------------------------------------------------

The Johannes Gutenberg University Mainz is currently constructing the
new continuous-wave multi-turn electron linac MESA (Mainz Energy
Recovering Superconducting Accelerator) on the Gutenberg
Campus~\cite{Hug2016}. For the \DM\ experiment, the high-power beam
dump of the accelerator will be used as a target for the possible
production of dark sector particles in the \si{\MeV} mass
range~\cite{Doria2018,Doria2019}. Once discovered, these could provide
information on the structure of dark matter, which makes up a large
proportion of our universe~\cite{Bjorken2009}.

The detector concept of the \DM\ experiment will implement
electromagnetic calorimeters surrounded by active veto counters. The
calorimeters will detect the transferred energy in elastic scattering
of the dark sector particles by atomic-shell
electrons~\cite{Christmann2019}, in which the energy range is defined
by the \SI{150}{\MeV} energy of the electron beam. The detector site
will be heavily shielded from the beam, blocking practically all
beam-related Standard Model particles. It is crucial for this
experiment that cosmogenic particles leading to background events are
vetoed with a high detection efficiency and homogeneity.

In nuclear and particle physics, the combination of
wavelength-shifting (WLS) fibers with a silicon photomultiplier (SiPM)
readout is known to be a viable option for the operation of a
scintillation counter since the first proofs of concept, {\em e.g.}\
in Refs.~\cite{Balagura2006,Andreev2005}, and the pioneering
applications in large numbers by the CALICE~\cite{CALICE2010} and T2K
Collaborations~\cite{Yokoyama2010,T2K2011}.  Today, it is an efficient
and robust technique that has found a widespread use in the
construction of tracking detectors and calorimeters, see for instance
the recent review on this branch of instrumentation in
Ref.~\cite{Simon2019}. A number of studies exist on the photoelectron
yields for different geometries and setups of scintillation counters
with embedded WLS fibers, see {\em e.g.}\
Refs.~\cite{Denisov2017,Artikov2018}.

The \DM\ experiment requires a highly efficient, compact, and
cost-effective veto system which is capable of muon identification and
also accommodates suppression of gamma-ray and neutron background.
The planned veto detector system will consist
%! in the order of
of approximately %!
80 box-shaped plastic scintillating counters, each of \SI{2}{\cm}
thickness and having a maximum size of \SI{25 x 200}{\square\cm}. The
counters will be arranged in two layers, and read out by SiPMs, the
latter possibly connected to WLS fibers for an enhanced light
collection.  Sheets of lead, between the inner and the outer veto
layer, should prevent low-energy $\gamma$-rays from reaching the
calorimeter.  In a Monte Carlo simulation using \texttt{Geant4} and
its internal optical transport routine, such scintillation counters
without WLS fibers showed a light yield of \SI{15}{photoelectrons}
(\pe) per \si{\MeV} energy deposition, also for the longest detector
element of \SI{200}{\cm} length. The light attenuation losses in the
simulation were below \SI{0.5}{\percent/\cm}.  For the full \DM\
experiment, the simulation predicts a veto efficiency of more than
\SI{99.9}{\percent} when employing this design.  For the study of the
remaining backgrounds at \DM\ a preliminary concept exists based on
(1) the analysis of beam-off data and (2) beam measurements with
rotated detector geometries.  This design for a veto system in the
search for dark matter at accelerators follows the approach of the BDX
Experiment at the Thomas Jefferson National Accelerator Facility
(JLab) in the USA~\cite{Battaglieri2018,Battaglieri2020}.

This paper describes studies of prototype counters for the \DM\ veto
system in the \SI[number-unit-product = -]{855}{\MeV} electron beam of
the Mainz Microtron (MAMI). A scintillation counter, in which
different configurations of fibers were embedded, is described in
Section~\ref{sec:setup}, the electron beam tests are presented in
Section~\ref{sec:beamtest}, the calibrations in
Section~\ref{sec:calibrations}, the light collection is discussed and
modeled as a function of the transverse distance from the beam
position to the fibers in Section~\ref{sec:modeling}, and the
conclusions are given in Section~\ref{sec:conclusions}.

%--------------------------------------------------------
\section{Description of the scintillation counters}
%--------------------------------------------------------
\label{sec:setup}

%--------------------------------------------------------
\begin{figure}[htbp]
  \centering
  \includegraphics[width=\columnwidth]{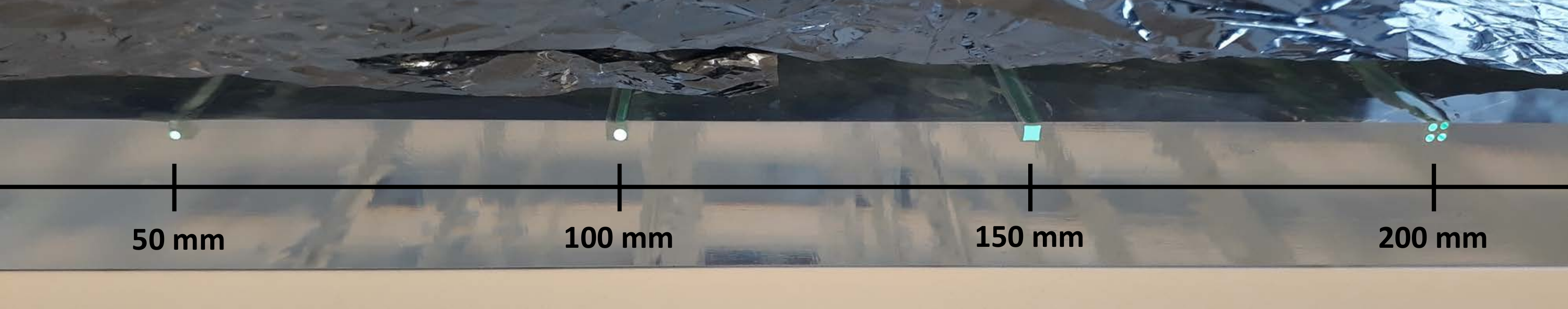}
  \caption{Photograph of one of the two read-out ends of the studied
    scintillation counter with fibers after polishing. Configurations
    from left to right: round fiber of
    \SI[number-unit-product=-]{1}{\mm} diameter, round fiber of
    \SI[number-unit-product=-]{1.5}{\mm} diameter, square fiber of
    \SI[number-unit-product=-]{2}{\mm} edge length, $2 \times 2$
    matrix of four round fibers of \SI[number-unit-product=-]{1}{\mm}
    diameter each.}
  \label{fig:ScintillatorPhoto}
\end{figure}
%--------------------------------------------------------

%--------------------------------------------------------
\begin{figure}[htbp]
  \centering
  \includegraphics[width=0.85\columnwidth]{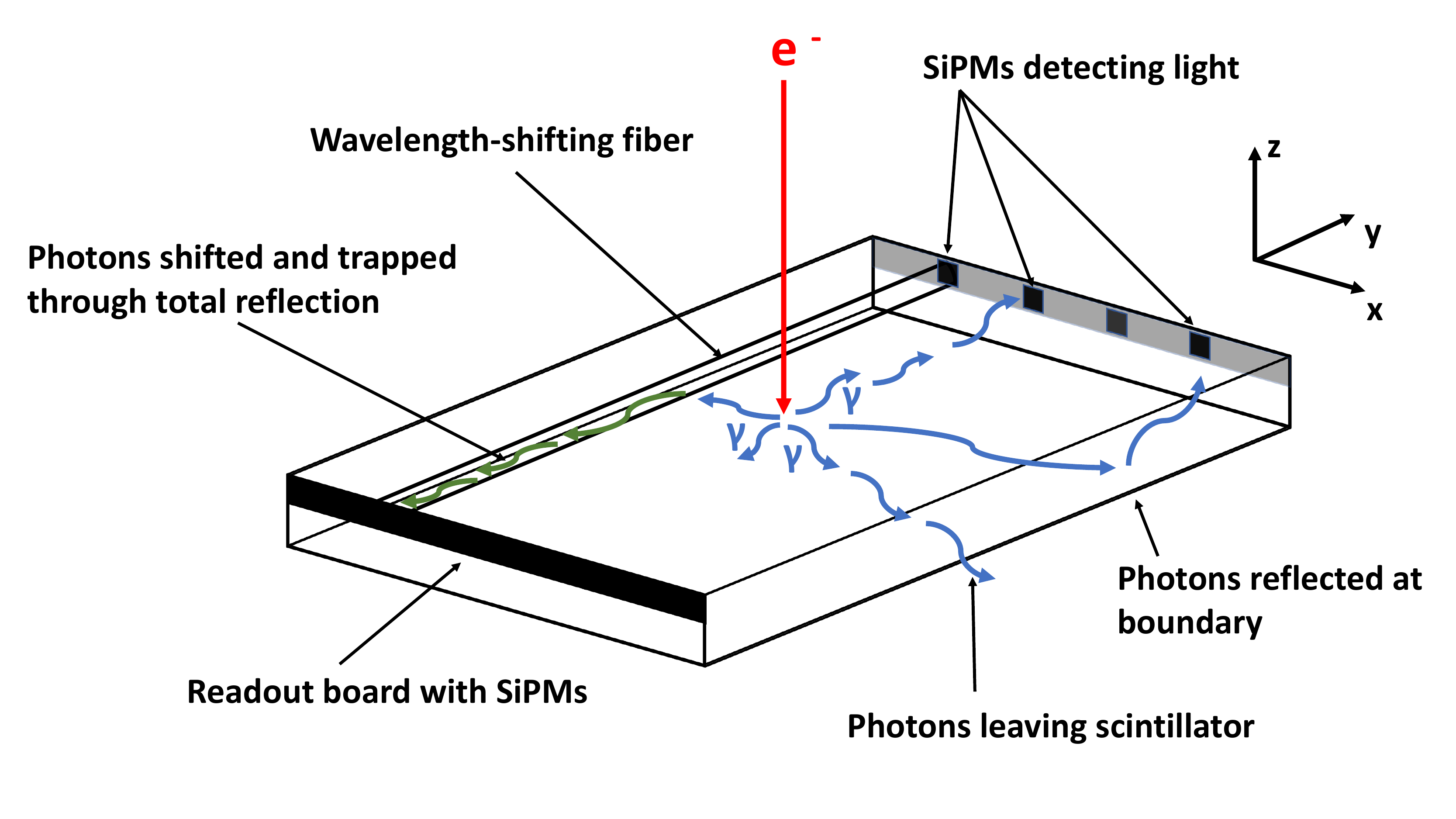}
  \caption{Schematic view of how a fiber is influencing the collection
    of scintillation light that is produced by a minimum ionizing
    electron beam penetrating the active volume of a counter. The
    scintillation counters in this study had a size of \SI{50 x 25 x
      2}{\cubic\cm}. The light was read out using four SiPMs mounted
    on a readout board on each of the two opposing ends.}
  \label{fig:ReadoutPrinciple}
\end{figure}
%--------------------------------------------------------

%--------------------------------------------------------
\begin{figure}[htbp]
  \centering \subfigure[Top
  layer]{\includegraphics[width=\columnwidth]{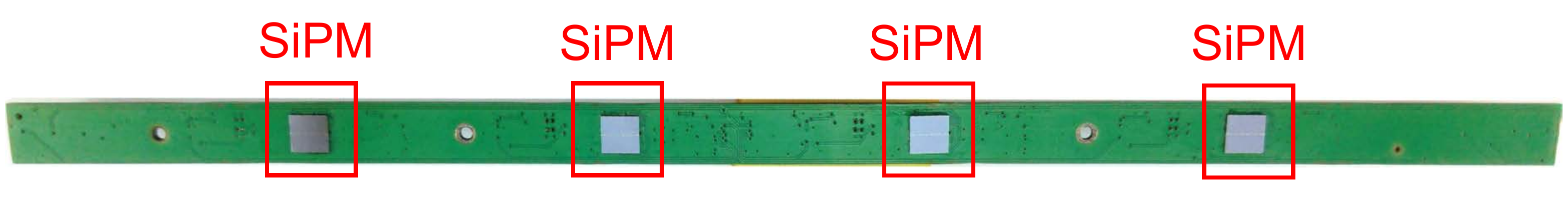}}
  \qquad \subfigure[Bottom
  layer]{\includegraphics[width=\columnwidth]{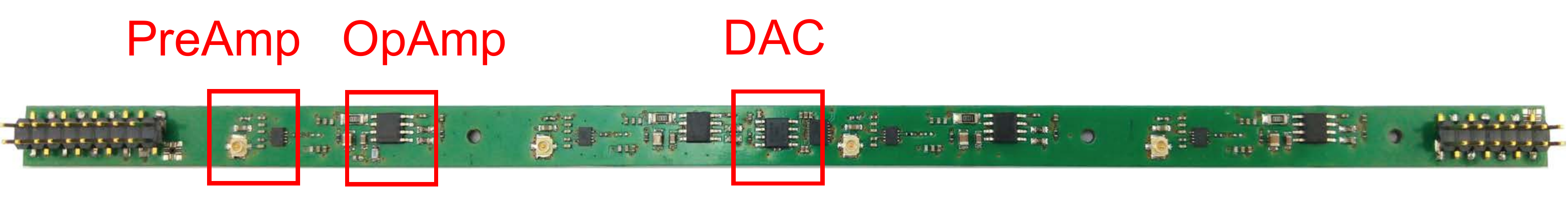}}
  \subfigure[Electronic circuit
  diagram]{\includegraphics[width=\columnwidth]{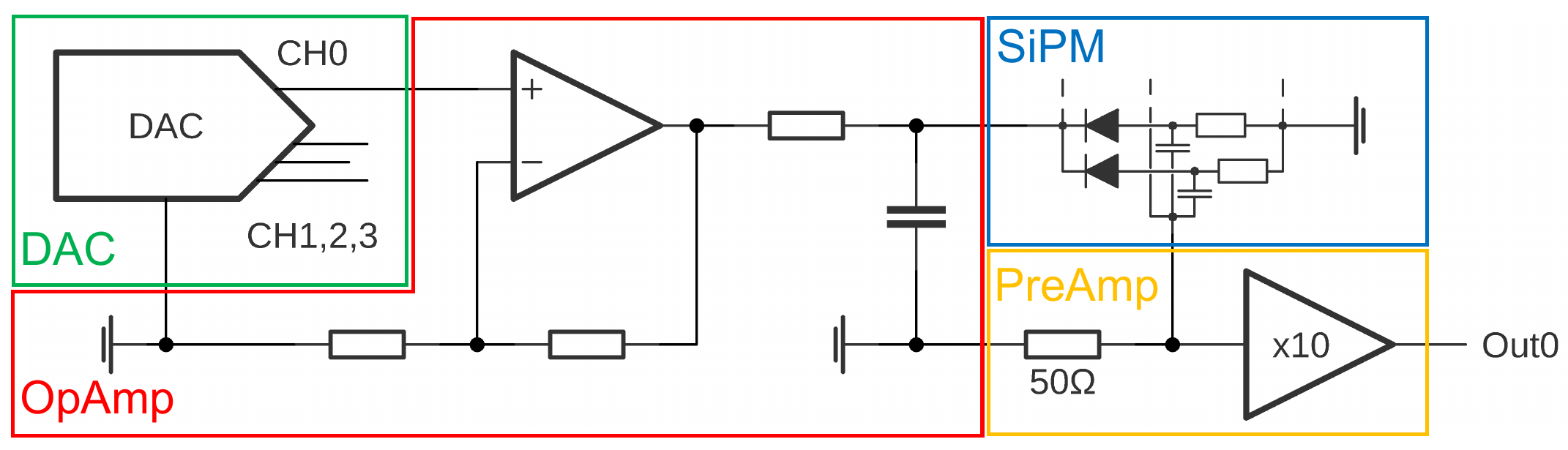}}
  \caption{Photographs of the top and bottom layer of the readout
    board and the electronic circuit diagram. The board has a size of
    \SI{1 x 25}{\square\cm}. (a) Top layer. Four SiPMs with \SI{6 x
      6}{\square\mm} active area, and a capacitively coupled fast
    output. (b) Bottom layer. PreAmp (one per SiPM): Signal
    preamplifier based on the gain block AD8354 with a transimpedance
    gain of $Z =$ \SI{500}{\ohm} and high analog bandwidth; OpAmp (one
    per SiPM): Non-inverting high-voltage operational amplifier
    circuit with current-limiting resistor for generating the bias
    voltage from an adjustable reference voltage; DAC (one per board):
    Digital-to-analog converter for setting the individual values for
    the reference voltages and thus the bias voltages. (c) Electronic
    circuit diagram.}
  \label{fig:SiPM}
\end{figure}
%--------------------------------------------------------

Two identical scintillation counters of type EJ-200 from Eljen
Technology~\cite{EJ200} were used in this study.  They had a size of
\SI{50 x 25 x 2}{\cubic\cm} and highly polished surfaces to promote
total internal reflection. The opposite ends of the counters were each
read out by four independent \SI{6 x 6}{\square\mm} SiPMs of type
J-Series 60035 from SensL~\cite{SensL}. The active area of the SiPMs
provide a sufficiently large coverage of the read-out ends of the
counter without fibers.  Parallel grooves of \SI{2.5 x
  2.5}{\square\mm} cross section were milled into the surface of one
counter, so that fibers could be placed into these grooves, which were
then filled with optical cement of type EJ-500 from Eljen
Technology. The protruding ends of the fibers were cut off and the two
readout sides of the counter were polished a second time. A photograph
of one of the finished read-out ends with the fibers can be seen in
Fig.~\ref{fig:ScintillatorPhoto}. The fibers were read out with the
same type of SiPM. For this counter, a smaller size of the SiPM active
area would have decreased noise as well as the detection of light not
coming from the fiber, but would have also complicated the precise
alignment to the groove and the comparison with the reference counter.

Four different fiber configurations were realized:
\begin{description}
\item[Ch0] Round WLS fiber of \SI[number-unit-product=-]{1}{\mm}
  diameter of type BCF-92 from Saint Gobain
  Crystals~\cite{SaintGobain}
\item[Ch1] Round WLS fiber of \SI[number-unit-product=-]{1.5}{\mm}
  diameter of type Y-11 (200) MJ from Kuraray~\cite{Kuraray}
\item[Ch2] Square scintillating fiber of
  \SI[number-unit-product=-]{2}{\mm} edge length of type BCF-20 from
  Saint Gobain Crystals~\cite{SaintGobain}
\item[Ch3] $2 \times 2$ matrix of four round WLS fibers of
  \SI[number-unit-product=-]{1}{\mm} diameter each, bundled together,
  of type BCF-92 from Saint Gobain Crystals~\cite{SaintGobain}
\end{description}
The fibers from Kuraray are based on a polystyrene core with a
refractive index of $n =$ \num{1.59}, while Saint Gobain specifies for
the core material $n =$ \num{1.60}. All three different fibers have a
thin inner cladding made from polymethylmethacrylate (PMMA) with $n =$
\num{1.49}. The efficiency for trapping emitted light within the core
of the round fibers ranges from $\epsilon =$ \SI{3.4}{\percent} for
light originating at the fiber axis to $\epsilon \sim$
\SI{7}{\percent} for light originating near the core--cladding
interface. For square fibers, $\epsilon =$ \SI{4}{\percent},
independent of the point of origin. All fibers had a second outer
cladding with $n = 1.42$, increasing the core trapping efficiencies by
about a factor of \num{1.6}. The light emission in the scintillating
fiber of channel Ch3 can also be excited by the blue light of the
scintillation counter, so that it can act analogous to a WLS fiber.
The peak emission wavelength of the different fibers is relatively
similar: \SI{476}{nm} for Y11 and \SI{492}{nm} for BCF-92 and
BCF-20. Based on the comparable light collection properties, the
differences between the fibers can only marginally influence the light
collection as a function of the transverse distance from the beam
position to the fibers.

A schematic view of the readout concept for the scintillation light is
presented in Fig.~\ref{fig:ReadoutPrinciple}. On the readout board
four individual SiPMs were powered each by an operational amplifier
generating the bias voltage from an adjustable reference
voltage. These voltages could be set by a digital-to-analog converter
through SPI protocol by the slow control software. The SiPMs from
SensL~\cite{SensL} feature a fast output signal with a significantly
smaller capacitance than the anode output and an intrinsic FWHM pulse
width of \SI{3}{\nano\s}, allowing for precision timing and high count
rates. All SiPMs were operated at a fixed bias voltage of
$V_\text{bias} =$ \SI{27.3}{\volt}, corresponding to an overvoltage of
$V_\text{OV} \sim$ \SI{3}{\volt} and leading to a typical gain of
$3 \times 10^6$. Note, that the temperature coefficient of the
breakdown voltage is $dV_\text{BD}/dT =$
\SI{21.5}{\milli\volt\per\kelvin}. The dark count rate of a single
\SI{6 x 6}{\square\mm} SiPM is about
\SI{64}{\kilo\hertz\per\square\mm} at $V_\text{OV} =$
\SI{3}{\volt}. The exact active area is \SI{6.07 x 6.07}{\square\mm},
the active area of each microcell is \SI{35 x
  35}{\square\micro\meter}, and there are 22\,292 microcells in a
sensor. The quoted fill factor is \SI{75}{\percent}. The
%! SiMPs
SiPMs %!
feature a photon detection efficiency (PDE) of \SI{40}{\percent}, an
afterpulsing probability of \SI{1}{\percent}, and a cross-talk
probability of \SI{10}{\percent} at $V_\text{OV} =$ \SI{3}{\volt}. The
outputs were connected to a signal preamplifier based on the gain
block AD8354 with a transimpedance gain of $Z =$ \SI{500}{\ohm} and
high analog bandwidth, leading to an internal amplification factor of
$G_{\text{int}} =$ \num{10} at $Z =$ \SI{50}{\ohm} input
impedance. The output signals were split into a timing branch and a
charge collection branch.  The trigger signal for the data acquisition
was realized with external electronic modules by discriminating the
analog sum of the signals in the timing branch. The trigger threshold
was set to \SI{30}{\milli\volt}, corresponding to \SI{15}{\percent} of
the most probable signal amplitude of a fully penetrating
minimum-ionizing particle.  The individual charge signals were
externally amplified by another factor of $G_{\text{ext}} =$ 10. The
signals of the SiPMs had a typical FWHM of \SI{14}{\nano\s} and a
width of \SI{30}{\nano\s} at the base, and were integrated during a
\SI{70}{\nano\s} long gate by a charge-sensitive ADC of type 2249A
from LeCroy with a sensitivity of \SI{0.25}{\pico\coulomb} per ADC
channel.

The readout board with the SiPMs and the front-end electronics are
shown in Fig.~\ref{fig:SiPM}. These boards were pushed onto the ends
of the scintillation counters by a mechanical support. Optical grease
was used to ensure optimal coupling of the SiPMs with these ends. The
active area of the SiPMs of \SI{6 x 6}{\square\mm} covered the grooves
of \SI{2.5 x 2.5}{\square\mm}. The remaining area of the SiPMs was not
optically shielded from the scintillator.  To increase the collection
of light, the scintillation counter was wrapped with aluminum-coated
Mylar foil on all sides except the read-out sides.

%--------------------------------------------------------
\section{Electron beam tests of the scintillation counters}
%--------------------------------------------------------
\label{sec:beamtest}

%--------------------------------------------------------
\begin{figure}[htbp]
  \centering
  \includegraphics[width=\columnwidth]{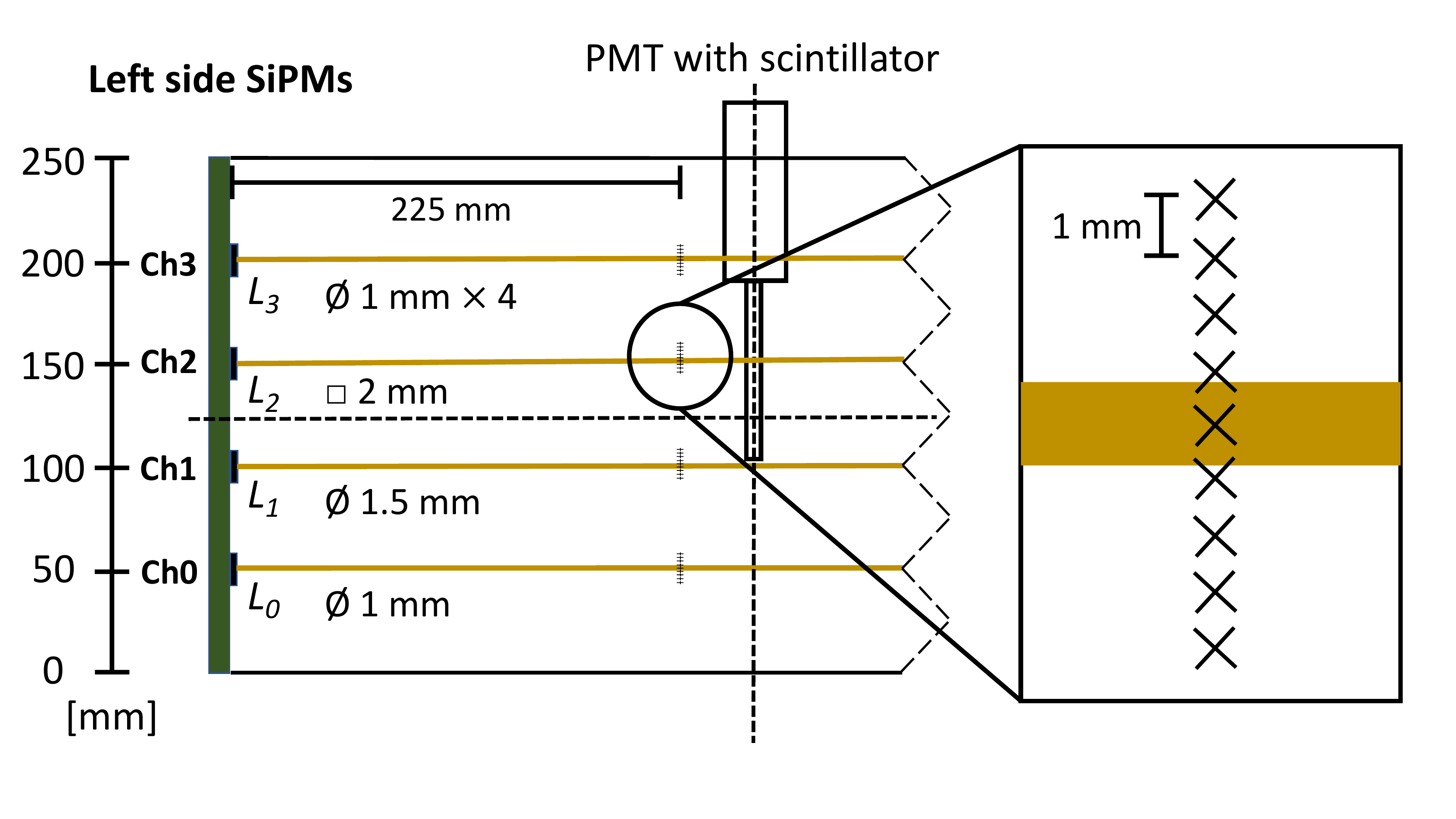}
  \caption{Positions of the electron beam (black crosses) on the
    \SI[number-unit-product=-]{250}{\mm} wide scintillation counter
    with fibers separated by \SI{50}{\mm}. Near each fiber nine
    positions with a pitch interval of \SI{1}{\mm} were scanned at a
    distance of \SI{225}{\mm} from the read-out side. The two dashed,
    perpendicular lines indicate the symmetry axes of the counter. The
    SiPMs of the four channels Ch0 to Ch3 at the left side are labeled
    $L_{0}$ to $L_{3}$. On the right, the relative scanning positions
    with respect to the fiber are shown in the enlarged view. The
    position of a separate, \SI{5}{\mm}-wide scintillator in the
    center position of the counter is indicated.}
  \label{fig:ScanWithFibers}
\end{figure}
%--------------------------------------------------------

%--------------------------------------------------------
\begin{figure}[htbp]
  \centering
  \includegraphics[width=\columnwidth]{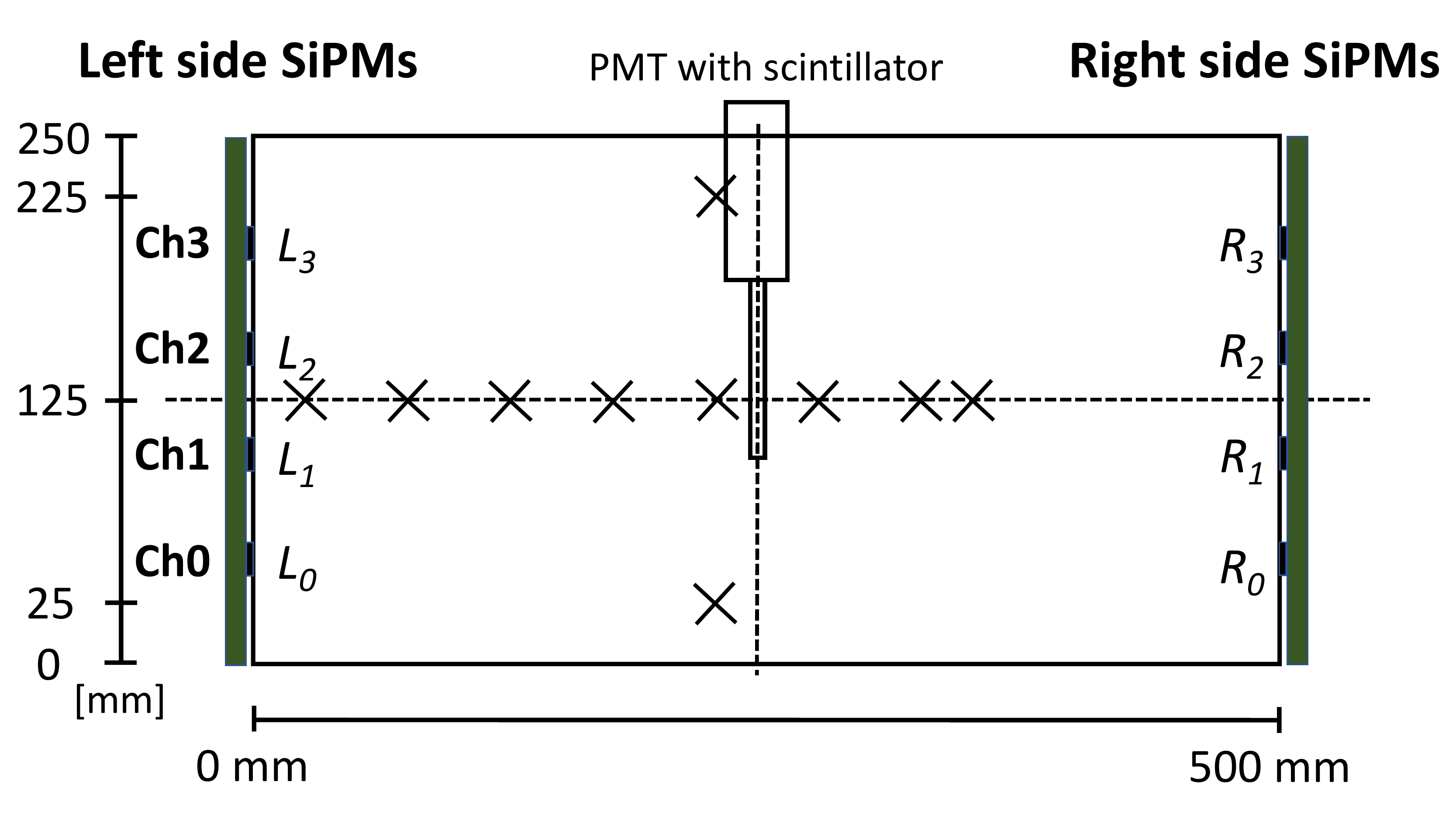}
  \caption{Positions of the electron beam (black crosses) on the
    \SI[number-unit-product=-]{250}{\mm} wide scintillation counter
    without fibers for the reference measurements. Three positions
    with a pitch interval of \SI{100}{\mm} were scanned at a distance
    of \SI{225}{\mm} from the read-out side. Seven additional
    measurements were taken along the central longitudinal axis. The
    two dashed, perpendicular lines indicate the symmetry axes of the
    counter. The SiPMs of the four channels Ch0 to Ch3 at the left and
    the right side are labeled $L_{0}$ to $L_{3}$, respectively
    $R_{0}$ to $R_{3}$. The position of a separate, \SI{5}{\mm}-wide
    scintillator in the center position of the counter is indicated.
  }
  \label{fig:ScanWithoutFibers}
\end{figure}
%--------------------------------------------------------

Electron beams of \SI{855}{\MeV} energy and sub-picoampere currents
from the Mainz Microtron MAMI were precisely pointed to a set of
positions on the top of one of the scintillation counters. The beam
electrons deposited $\Delta E \approx$ \SI{4}{\MeV} energy in the
scintillator bulk, leading to $\mathcal L \sim$ 40\,000 scintillation
photons at the site of the ionizations.  The typical analog sum of all
eight SiPM output signals had amplitudes of \SI{200}{\milli\volt}. The
environmental temperature in the experiment was stabilized to minimize
voltage drifts. Noise amplitudes were well below the trigger threshold
and the trigger efficiency was estimated to be very close to
\SI{100}{\percent}.  The measurements had a duration of \SI{60}{\s} at
a data acquisition rate of \SI{2}{\kilo\Hz} resulting in more than
$10^3$ events in each charge spectrum. At these trigger rates,
cosmogenic events and beam-unrelated background in the data set could
be safely neglected.

The detector was placed in a dark box to shield it from external light
sources and the whole setup was supported by a remotely steerable
$x$-$y$ table.  The beam left the vacuum beam pipe through an aluminum
flange of $\Delta z \sim$ \SI{0.2}{\mm} thickness and traversed
$z \sim$ \SI{500}{mm} of air before hitting the detectors. Multiple
scattering in the flange increased the divergence of the beam to
$\theta_{\text{beam}} <$ \SI{1}{\milli\radian}, resulting in a beam
spot of $\sigma_{\text{beam}} \ll$ \SI{1}{\mm} width at the position
of the counters.  The beam position along the longitudinal axis was
determined by a \SI{5}{\mm}-wide, separate scintillation detector
coupled to a photomultiplier tube and located at the center position
of the studied counters.  The precision for the position with respect
to the beam can be conservatively estimated with the detector width.
The position on the transverse axis was determined by scanning the
counter to its upper edge. This allowed to determine its position with
respect to the beam with a precision of $\delta x \le$ \SI{1}{\mm}.
    
To study the light collection of the different configurations as a
function of the transverse distance from the electron beam to a fiber,
a scan parallel to the read-out side of the counter was performed as
depicted in Fig.~\ref{fig:ScanWithFibers}. The beam position relative
to the fibers was determined from the scans with sub-millimeter
precision.  For reference, corresponding measurements were performed
with the scintillation counter without embedded fibers. Three
positions of the electron beam at the same distance from the read-out
side were scanned as seen in Fig.~\ref{fig:ScanWithoutFibers}.

%--------------------------------------------------------
\section{Calibration of the charge spectra}
%--------------------------------------------------------
\label{sec:calibrations}

%--------------------------------------------------------
\subsection{Evaluation of the number of photoelectrons}
%--------------------------------------------------------

Calibrations were performed in a separate study immediately after the
beam tests under the same environmental conditions, in particular the
same temperature. To convert the ADC values into a number of \si{\pe},
each SiPM was exposed to short LED light pulses at a wavelength of
\SI{450}{\nano\m}. The custom-made LED and pulser setup featured a
small pulse-to-pulse variation. The LED light was optically diffused
\SI{40}{\cm} in front of the SiPMs and guaranteed a Poisson
distributed number of photons per pulse sufficiently large to be in
the Gaussian limit. Consequently, the resulting charge spectra showed
symmetric, Gaussian distributed peaks with relative widths of the
order of \SI{10}{\percent}. If one assumes that the width of such a
peak is caused by statistical fluctuations only, it will follow that
$\sigma/\hat{n} = \sqrt{\lambda}/\lambda = 1/\sqrt{\lambda}$\,, where
$\hat{n}$ is representing the position of the peak maximum, $\sigma$
the peak width, and $\lambda$ being the mean and variance of the
Poisson distribution for the number of \si{\pe} The electronic noise
was determined independently from the LED by a pulser with a
repetition frequency of \SI{80}{\Hz} and was significantly smaller
than the peak widths.  Including the subtraction of the measured
pedestals in the charge spectra leads to the relation:
\begin{equation}
  % \label{eq:PE}
  \frac{\sqrt{\sigma_{\text{LED}}^{2} - \sigma_{\text{ped}}^{2} } }
  {\hat{n}_{\text{LED}} - \hat{n}_{\text{ped}} } = \frac{1}{\sqrt{\lambda}}\,,
\end{equation}
where $\hat{n}_{\text{ped}}$ and $\sigma_{\text{ped}}$ are the
position and width of a fit to the pedestal peak with a Gaussian
distribution.

The conversion factors $c_{i}$ of calibrated ADC channels per \si{\pe}
were determined for each SiPM at a wide range of bias voltages from
the calibrated peak positions
$c_{i} = (\hat{n}_\text{LED} - \hat{n}_\text{ped}) \cdot \kappa_{i} /
\lambda$. The damping factors $\kappa_{i}$ needed to be included for
each one of the eight SiPMs to account for signal losses through the
cable pathways. They were determined by sending a well-defined amount
of charge in pulses of a high precision frequency generator of type
81160A from Keysights Technologies through the signal pathways to the
ADC.

The breakdown voltages $V_\text{BD}$ of each SiPM were determined from
this series of measurements by the projection of the measured gain to
zero. The spread in breakdown voltages was up to \SI{0.5}{\volt}, in
good agreement with the range specified in the data sheet.

%--------------------------------------------------------
\subsection{Evaluation of the light yield}
%--------------------------------------------------------

%--------------------------------------------------------
\begin{figure}[htbp]
  \centering \includegraphics[width=\columnwidth]{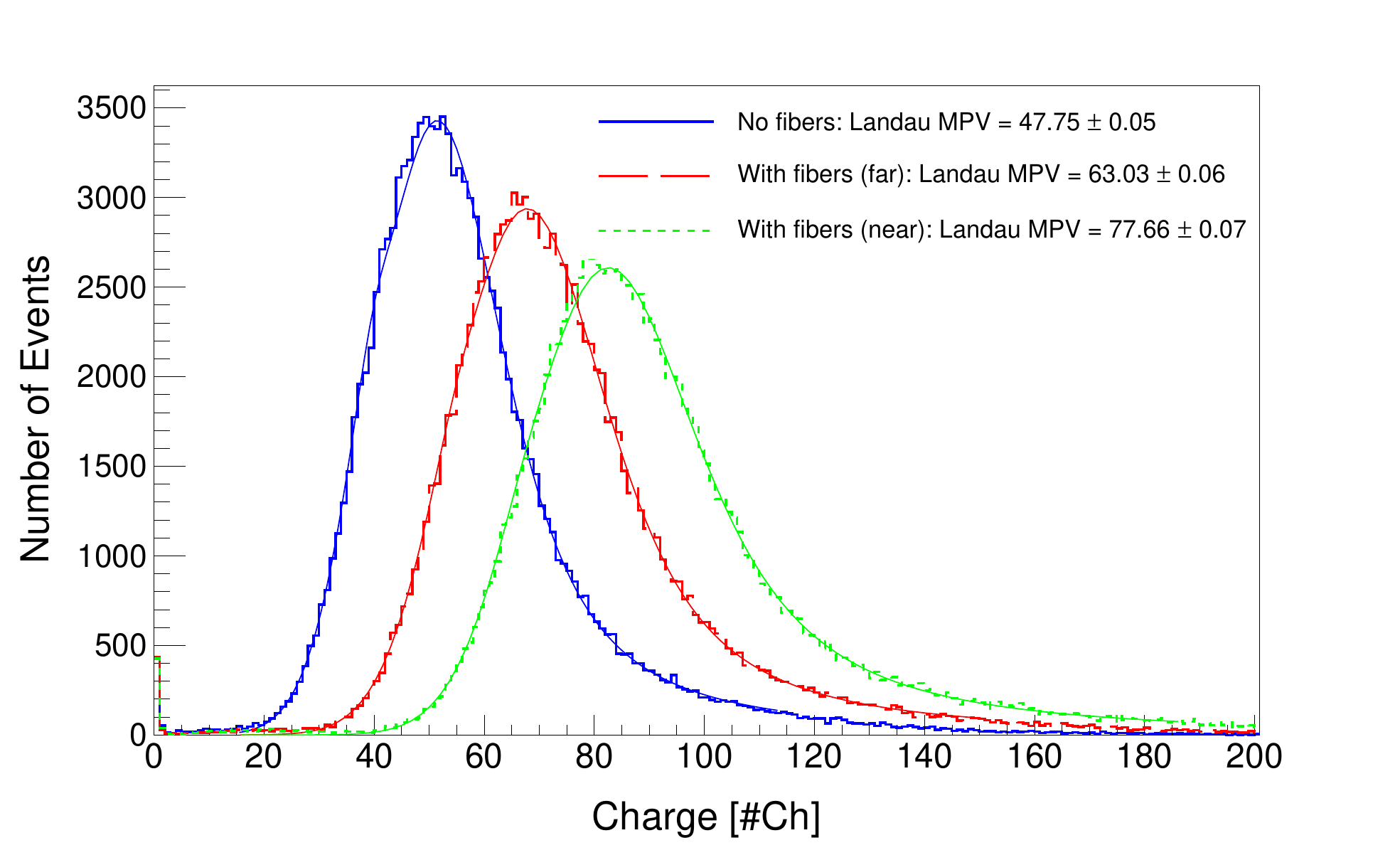}
  \caption{Typical ADC spectra ($\#\text{Ch} \mathrel{\widehat{=}}$
    \SI{0.25}{\pico\coulomb}) without calibration for a single SiPM
    recorded when the minimum ionizing electron beam penetrated the
    counter in a distance of \SI{225}{\mm} from the read-out side. For
    the left spectrum, a scintillation counter without embedded fibers
    was used. The most probable value of this distribution corresponds
    to a light yield of $\lambda \simeq$ \SI{25}{\pe} after
    calibration. The center and right spectra were taken with a fiber
    of round geometry with diameter of \SI{1.5}{\mm} in a transverse
    distance of \SI{4}{\mm} (near) and \SI{104}{\mm} (far) to the beam
    position, respectively.  The asymmetric peak shape could be well
    described by a Landau distribution convoluted with a Gaussian
    distribution.}
  \label{fig:ADCspectrumWithFit}
\end{figure}
%--------------------------------------------------------

Figure~\ref{fig:ADCspectrumWithFit} shows typical ADC spectra of a
single SiPM, recorded when the scintillation light was produced by the
electron beam penetrating the counter.  The asymmetric peak shape
could be well described by a Landau distribution to describe the
physical distribution of the energy loss, convoluted with a Gaussian
distribution with width parameter $\sigma$ to incorporate noise and
other smearing effects. The most probable value $\hat{n}_{\text{MPV}}$
of this function was used as a measure for the light yield, where the
mean number of \si{\pe} is given by
$\lambda = (\hat{n}_{\text{MPV}} - \hat{n}_{\text{ped}}) \cdot
\kappa_{i}/c_{i}\,$. The ADC pedestals were determined in separate
measurements.

%!
The light yield $\lambda$ will be given in units of \si{\pe} in this
paper. A conversion to the number of incident photons $Y$ in units of
\si{\ph} is wavelength-dependent and may imply additional
uncertainties.  The number of discharged microcells in a SiPM for a
light pulse of moderate intensity is proportional to its PDE. It
depends further on the sum $q$ of the correlated noise of cross-talk
and afterpulsing in the time window of the gate, leading to
$Y = \frac{ 1 - q }{\mathit{PDE} }\, \lambda$. The individual
conversion factors for each SiPM were determined by a linear
interpolation of the data sheet values for the PDE and the cross-talk
plus afterpulsing probabilities to $V_\text{OV} \sim$
\SI{3}{\volt}. They were typically on the order of
\SI{2.4}{\ph\per\pe} for the blue light of the scintillation counters
and \SI{20}{\percent} smaller for the light emitted by the fibers,
with a larger reduction for the BCF fibers than for the Kuraray fiber.
% !

%--------------------------------------------------------
\section{Analysis and modeling of the light yield}
%--------------------------------------------------------
\label{sec:modeling}

%-------------------------------------------------------
\subsection{Reference light yield from a counter without fibers}
%-------------------------------------------------------

For a counter without fibers, the measured light yield was almost
constant for beam positions along the transverse axis: the four inner
SiPMs (Ch1 and Ch2, left and right) showed a variation of less than
\SI{1}{\percent}, while the four outer SiPMs (Ch0 and Ch3, left and
right) showed a decrease or increase of not more than
\SI{3}{\percent}. These observations could be explained by light being
produced in a thin counter that will get distributed almost
homogeneously over the volume due to the many internal reflections.
The average over the channels and transverse positions of the most
probable light yields observed on the individual SiPMs was determined
to be $\lambda_{\text{ref}} =$ \SI{23.0 +- 0.2}{\pe} and then used as
a reference value for the light yield for a counter without fibers.

The effective attenuation of the light produced along the central
longitudinal axis was determined by the eight measurements indicated
in Fig.~\ref{fig:ScanWithoutFibers}. For beam positions at distances
of more than \SI{17}{cm} from the read-out side, no significant
difference in light yield between the four channels of either side was
found.  At these positions the sum of the light yield from the
corresponding two SiPMs on both sides was almost constant.  The
observed attenuation in longitudinal direction was less than
\SI{0.5}{\percent/\cm}. Single exponential fits to the light yield at
more than \SI{17}{cm} from the read-out sides resulted in an average
effective light attenuation length of $\Lambda_{\text{att}} =$
\SI{220}{\cm}.  The light sharing between the two sides, expressed as
an asymmetry $\mathcal{A} = (L_k-R_k)/(L_k+R_k)$ for a channel Ch$k$,
allowed to determine the longitudinal position with a precision of
$\delta z \sim$ \SI{25}{mm}.

%-------------------------------------------------------
\subsection{Light yield from a counter with fibers}
%-------------------------------------------------------

%--------------------------------------------------------
\begin{figure}[htbp]
  \centering \subfigure[Left end
  SiPMs]{\includegraphics[width=\columnwidth]{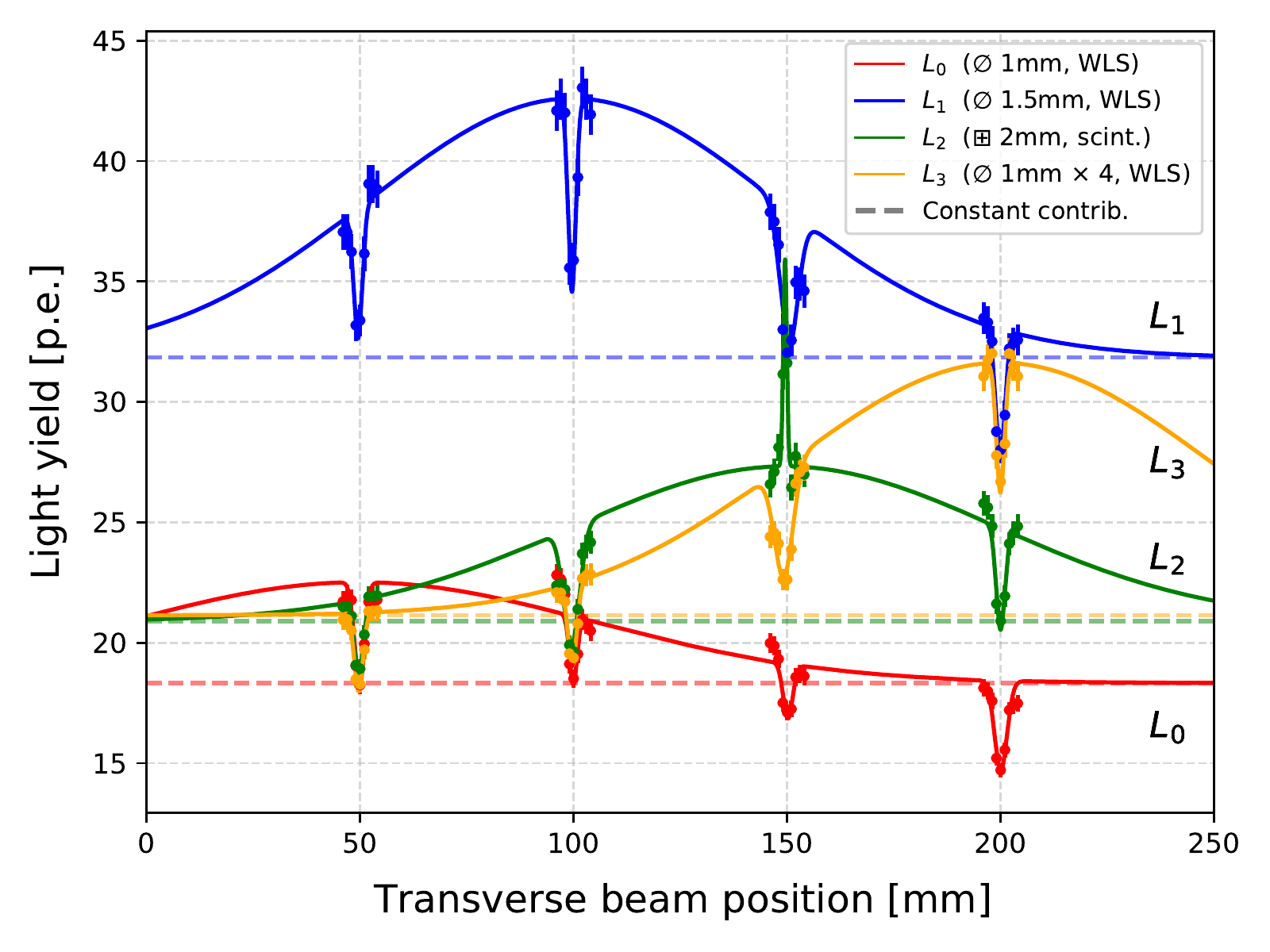}} \qquad
  \subfigure[Right end
  SiPMs]{\includegraphics[width=\columnwidth]{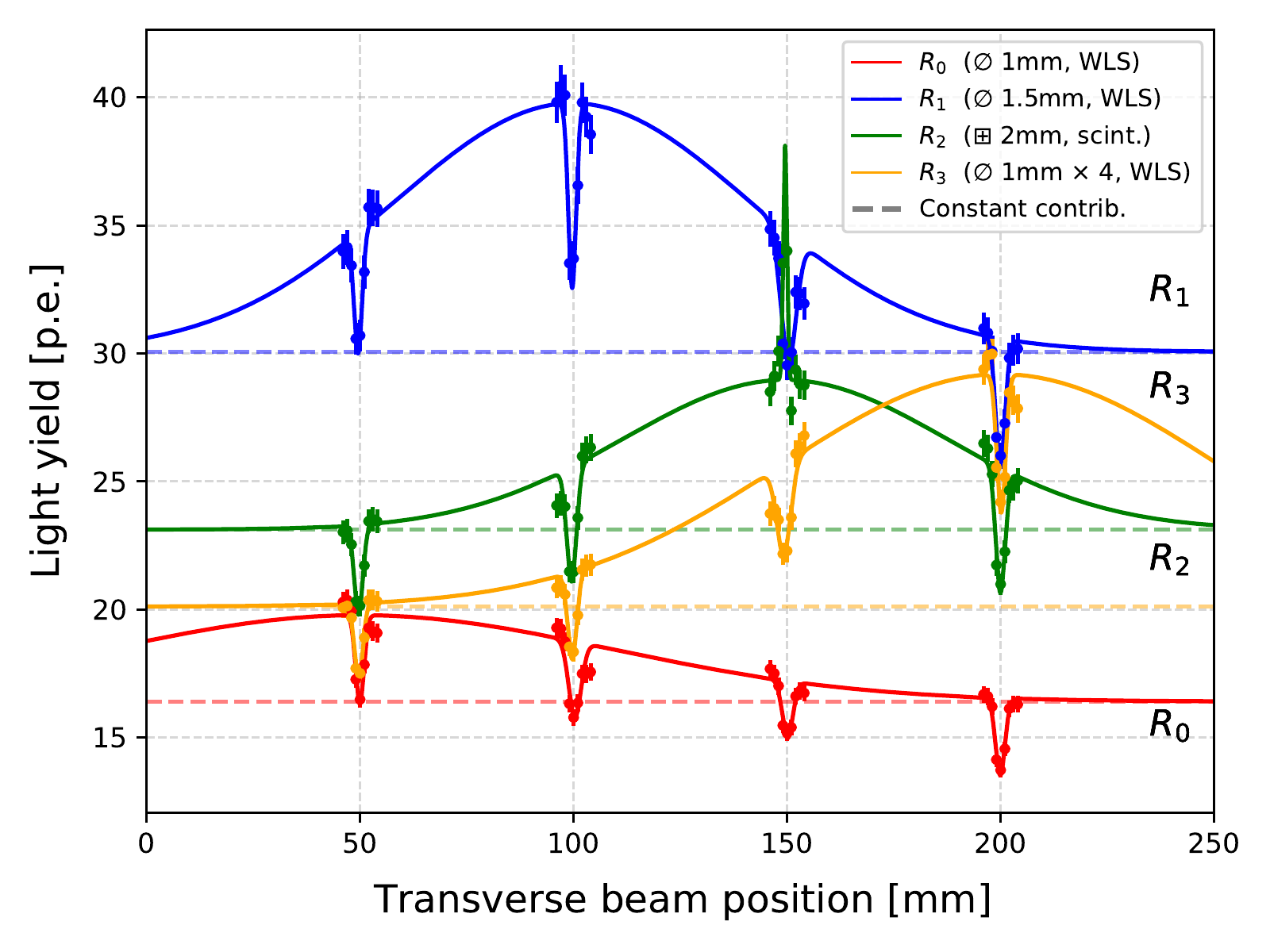}}
  \caption{Light yield in number of photoelectrons for each SiPM
    connected to a fiber as a function of the transverse position of
    the beam.  (a) Left end SiPMs. (b) Right end SiPMs.  A broad peak
    on top of a wide distribution can be observed, with the maximum of
    the peak at the respective fiber position, where the beam was
    located.  The effect of the grooves on the light yield is visible.
    The curves show a model description for the total light yield
    (full line), consisting of a constant (dashed line), local
    structures at the groove positions, and a peaking contribution.  }
  \label{fig:LightYieldFibers_global}
\end{figure}
%--------------------------------------------------------

Figure~\ref{fig:LightYieldFibers_global} shows the light yield from a
counter with fibers as a function of the transverse position of the
beam, measured at a longitudinal distance of \SI{225}{\cm}.

For all channels, the light yield across the transverse width of the
counter showed a broad peak on top of a wide distribution, with the
maximum position of the peak at the respective fiber position. As each
fiber was placed in a groove of \SI{2.5}{\mm} depth, the reduced
thickness of the scintillating material implied a local decrease in
light yield at these positions. In case of the square fiber, the
expected scintillation of the fiber was observed for direct electron
beam exposure.

To model these observations, different descriptions of the wide
distribution, the broad peak, and the local structures at the groove
positions were tested.

It was found, that the wide distribution was best described by a
constant.  This is in agreement with the study of the counter without
fibers, where no significant variation in light yield was observed
between the four SiPMs on either side.

To describe the local structures at the groove positions, a sinc
function, the second derivative of a Gaussian function, and a Gaussian
function were tested. The nominal position of each fiber was
determined by taking the mean value of all eight extremal positions
from both ends, left and right, of the counter.

A well-motivated position dependence for the broad peak could be
derived from the solid angle coverage of a fiber for direct
light. This leads to the proportionality
$\Delta \Omega \propto 1/\sqrt(x^2 + z^2)$ with $x$ as transverse
distance of the fiber from the light source and $z$ as the depth of
the light source inside of the scintillator. After integration over
$z$ from 0 to the thickness of the scintillator, this function shows a
steep dependence close to the fiber. However, for larger distances,
this dependence becomes inconsistent with the data. At these
distances, a solid angle coverage approximated by
$\Delta \Omega \propto 1/(x^2+z^2)$, corresponding to an infinitely
long fiber, provides a better description.  A combination of both
approaches leads to two complications: (1) With additional parameters
the fit results are less robust and (2) the interpretation of the
interplay of several effects or approximations in a sophisticated
model becomes very difficult and less reliable. In contrast, a
Gaussian or a Lorentz distribution can describe the broad peak
reasonably well and provide at the same time (1) a very robust fit and
(2) have only few parameters, that all have a clear interpretation.
To conclude, some sophisticated models can provide a superior
description of the data points in certain ranges compared to simpler
functions, but fail in other ranges and add unnecessary complexity.

The best simultaneous fit to all data points from a channel Ch$k$ was
found with a linear combination of a constant yield
$\lambda_\text{far}^k$, four Gaussian distributions at the four groove
positions $\mu_{0-3}$, and one Gaussian distribution for the peaking
contribution $\lambda_\text{peak}^k$ at the respective fiber position
$\mu_{k}$:
\begin{equation}
  \lambda^{k}(x) = \lambda_\text{far}^k
  + \lambda_\text{peak}^k\cdot
  \text{e}^{ -\frac{ (x - \mu_{k})^2 }{ 2\sigma_{k}^2 }}
  + \sum_{i = 0}^{3} \lambda_\text{i}^k \cdot
  \text{e}^{ -\frac{ (x - \mu_{i})^2 }{ 2\sigma_{i}^2 }}
\end{equation}
The observed light yields for the beam position {\em near} and {\em
  far} from the fiber, and a broad {\em peak} in the distribution,
with
$\lambda_\text{near}^k = \lambda_\text{far}^k +
\lambda_\text{peak}^k$, motivated the naming of the parameters.  This
model resulted in a statistically acceptable $\chi^2$ for a reasonably
low number of fitted parameters, where the number of degrees of
freedom ({\it n.d.f.} $ = 21$) equals the number of scanned beam
positions minus the number of fitted parameters. The interesting
parameter values and the $\chi^2/\mathit{n.d.f.}$ as a measure for the
goodness of the fit are listed in Table~\ref{tab:lightyield}.

%--------------------------------------------------------
\subsection{Systematic uncertainties and bias}
%--------------------------------------------------------

Systematic uncertainties of the light yields between channels could
arise from the opto-mechanical coupling of the sensors with the
readout sides of the counter. Such channel-to-channel uncertainties
were estimated to be on the order of a few percent. Sources for shifts
up or down of all light yields leading to correlated systematic
uncertainties were not considered in this study.  Factors contributing
to a bias in the comparison of the light yield between channels could
be due to different fiber characteristics such as different trapping
efficiencies or attenuations in the fibers, their spectral matching
with the SiPMs, or small temperature variations leading to different
overvoltages of the SiPMs.

The systematic uncertainties of the light yields from the fitting and
calibration procedures within one channel were estimated with
\SIrange{1}{2}{\percent}. This range was evaluated from the variation
of the residua of different model predictions and data points as well
as the robustness of the fit results with respect to changes in the
data and the calibrations. In the further analysis, the quoted
uncertainties account for these point-to-point uncertainties within
one channel as these are of relevance in the context of this study.

%--------------------------------------------------------
\subsection{Interpretation of the results}
%--------------------------------------------------------

The total light yield of the counter with fibers can be interpreted as
composed of two contributions:
\begin{enumerate}
\item One contribution to the collected light has a very weak
  dependence on its point of origin.
\item The second contribution to the collected light has a strong and
  peaking dependence on its point of origin.
\end{enumerate}

The total light yield depends on many technical and geometrical
aspects, such as groove size, coverage of SiPM active area, etc.,
while the peaking contribution can be attributed purely to the fiber.

The first contribution could be explained in analogy to the case of
the counter without fibers: A fiber collects a certain fraction of the
indirect light, so that one contribution to the light yield from a
fiber should be varying only weakly with respect to the transverse
direction. Besides, the active area of the SiPMs outside of the groove
cross section is exposed to the scintillator bulk and adds to the
light yield. This contribution $\lambda_\text{bulk}$ was calculated by
scaling the measured reference light yield from the counter without
fibers by the uncovered SiPM active area in the counter with
fibers. The second contribution could be explained by light directly
emitted into the solid angle covered by a fiber. This contribution
should steeply increase as the position of the light production gets
closer to the fiber.

%--------------------------------------------------------
\begin{table*}[htbp]
  \centering
  \caption{Separated light yield contributions for a beam position far
    from the respective fiber $\lambda_\text{far}$, the broad peak
    $\lambda_\text{peak}$, and the sum of both with
    $\lambda_\text{near} = \lambda_\text{far} + \lambda_\text{peak}$,
    averaged for each pair of SiPMs from the left and the right side
    of a fiber. The last line gives the reference value
    $\lambda_\text{ref}$ from the counter without fibers.  The quoted
    uncertainties account for point-to-point uncertainties within one
    channel.\newline }
  \label{tab:lightyield}
  \begin{tabular}{lrrrcc}
    \toprule
    {Channel} & {$\lambda_\text{far}$ (\si{\pe})} & {$\lambda_\text{peak}$ (\si{\pe})} & {$\lambda_\text{near}$ (\si{\pe})} & \multicolumn{2}{S}{$\chi^{2}/${\it n.d.f.}} \\ 
              & & & & \text{Left} & \text{Right} \\
    \midrule
    Ch0 ($\varnothing$ \SI{1}{\mm}) & \num{17.0 +- 0.4} & \num{3.5 +- 0.5} & \num{20.5 +- 0.7} & \num{3.1} & \num{2.2} \\
    Ch1  ($\varnothing$ \SI{1.5}{\mm}) & \num{30.6 +- 0.5} & \num{10.1 +- 0.6} & \num{40.7 +- 0.8} & \num{1.2} & \num{1.4}\\
    Ch2 ($\boxplus$ \SI{2}{\mm}) & \num{22.8 +- 0.4} & \num{6.0 +- 0.4} & \num{28.8 +- 0.6} & \num{1.8} & \num{1.5} \\
    Ch3 ($\varnothing$ \SI{1}{\mm} $\times$ 4) & \num{20.7 +- 0.2} & \num{9.9 +- 0.4} & \num{30.6 +- 0.5} & \num{1.2} & \num{1.7} \\ 
    \midrule
    Reference (no fibers) & \num{23.0 +- 0.2} \\
    \bottomrule
    \end{tabular}
\end{table*}
%--------------------------------------------------------

% -------------------------------------------------------
\begin{table*}[htbp]
    \centering
    \caption{Gains in light yield
      $G_\text{far} = (\lambda_\text{far} -
      \lambda_\text{bulk})/\lambda_\text{ref}$,
      $G_\text{peak} = \lambda_\text{peak}/\lambda_\text{ref}$, and
      $G_\text{near} = G_\text{far} + G_\text{peak}$ averaged for each
      pair of SiPMs from the left and the right side of a fiber, with
      respect to the reference value from the counter without fibers.
      The correction for the partial coverage of the bulk scintillator
      by the SiPM active area has been taken into account for
      $G_\text{far}$, which could add a systematic bias.  The
      difference between $G_\text{near}$ and $G_\text{far}$ for one
      fiber configuration quantifies the transverse inhomogeneity of
      the light yield imposed by the fibers.  All gains would be
      larger by a factor of \num{5.75}, if the SiPMs were covering
      only the groove, see text. The quoted uncertainties account for
      point-to-point uncertainties within one channel. \newline }
    \label{tab:comparison}
    \begin{tabular}{lrrr}
    \toprule
    {Channel} & {$G_\text{far}$ (\si{\percent})} & {$G_\text{peak}$ (\si{\percent})} & {$G_\text{near}$ (\si{\percent})}\\ 
    \midrule
    Ch0 ($\varnothing$ \SI{1}{\mm}) & \num{-8.9 +- 1.7} & \num{15.0 +- 1.9} & \num{6.2 +- 2.6} \\
    Ch1 ($\varnothing$ \SI{1.5}{\mm)} & \num{50.3 +- 2.4} & \num{43.9 +- 2.3} & \num{94.2 +- 3.4} \\
    Ch2 ($\boxplus$ \SI{2}{\mm}) & \num{16.1 +- 1.7} & \num{26.0 +- 1.8} & \num{42.1 +- 2.5} \\
    Ch3 ($\varnothing$ \SI{1}{\mm} $\times$ 4) & \num{7.1 +- 1.2} & \num{42.8  +- 1.4}  & \num{49.9 +- 1.9} \\
    \bottomrule
    \end{tabular}
\end{table*}
% -------------------------------------------------------

Table~\ref{tab:comparison} shows the gains in light yield for the
separated contributions. All fiber configurations improved the light
yield, when the light was being produced close to the respective fiber
position. The gain with respect to the reference counter without
fibers and related to the direct light collection was quantified by
$G_\text{peak} = \lambda_\text{peak}/\lambda_\text{ref}$. It was found
to be in the order of \SI{45}{\percent} for the
\SI[number-unit-product=-]{1.5}{\mm} fiber and distinctively lower for
the other fiber configurations.  The gain with respect to the
reference counter without fibers for the indirect contribution was
analogously quantified by
$G_\text{far} = (\lambda_\text{far} -
\lambda_\text{bulk})/\lambda_\text{ref}$.  The
\SI[number-unit-product=-]{1.5}{\mm} fiber was improving the
collection of indirect light in the same order of magnitude as the
collection of direct light.  In the other configurations, the gain for
the indirect light collection was significantly lower.  An increase in
light yield by using fibers is invariably linked with a loss of
homogeneity of light collection in the transverse distance from the
point of origin to the fibers, when the light production is located in
sufficient distance from the read-out side. For completeness, the near
gain was defined as $G_\text{near} = G_\text{far} +
G_\text{peak}$. Then, the difference between $G_\text{near}$ and
$G_\text{far}$ for one fiber configuration quantifies the transverse
inhomogeneity of the light yield imposed by the fibers. If the gains
were calculated by interpolating the SiPM size to the groove size,
using
%! $G_\text{far}^\text{matched} = (Y_\text{far} -
%! Y_\text{bulk})/(Y_\text{ref} - Y_\text{bulk})$
$G_\text{far}^\text{matched} = (\lambda_\text{far} - %!
\lambda_\text{bulk})/(\lambda_\text{ref} - \lambda_\text{bulk})$ %!
and the analogous definition for $G_\text{peak}^\text{matched}$, all 
gain values would be larger by a factor of
$\lambda_\text{ref}/(\lambda_\text{ref} - \lambda_\text{bulk}) = $
\num{5.75}. Relative to each other all values keep their
proportion. Those gains provide a theoretical comparison of the light
yield from a scintillation counter with fibers to a reference counter,
where the SiPMs in both detectors have the size of the grooves,
leading to the maximum possible gain that can be achieved with a
well-chosen SiPM size. In practice, small-sized SiPMs are preferable
for the detector with fibers and large-sized SiPMs are preferable for
the detector without fibers.

% -------------------------------------------------------
\section{Conclusions}
%--------------------------------------------------------
\label{sec:conclusions}

This work has shown that, {\em e.g.}, a WLS fiber with a round
geometry and a diameter of \SI{1.5}{\mm} significantly increased the
light yield from a box-shaped counter of \SI{2}{\cm} thickness in
which it was embedded.  For configurations with fibers of smaller
diameter or made from a bundle of such fibers, the gain in light yield
was distinctively smaller.  It was observed that the gain increased
strongly for a point of origin of the light within distances of $\pm$
\SI{10}{cm} to the fiber.  The data were modeled by attributing the
sum of two contributions to the total light yield: the collection of
indirect light and of direct light reaching a fiber.  Based on the
observation, one should place this type of fibers in veto counters
such as the one for \DM\ with a maximum distance of $\SI{10}{cm}$ to
profit from the collection of direct light.

The dependence of the light yield on the point of origin of the light
can lead to complications in the interpretation of the SiPM output
signals in WLS fiber configurations. For example, in the BDX
Experiment at JLab, a detailed description of the counter geometry and
the photoelectron response needed to be implemented in a simulation
framework to account for these
complications~\cite{Battaglieri2020}. On the other hand, the
transverse position sensitivity could have a positive effect, for
instance to determine the point of origin within the counter with an
increased resolution when considering signal intensities of multiple
SiPMs. On the contrary, the longitudinal position resolution provided
by the light sharing in a counter without fibers could also be an
advantage.

The veto counters for the small \DM\ prototype, that has a size of
\SI{50 x 50 x 80}{\cubic\cm}, were constructed without embedded
fibers, retaining the relative ease of construction of the veto
system. The design of the readout board was optimized for this
application and incorporates nine instead of four SiPMs, thereby
increasing the total light yield and improving the uniformity at the
read-out ends.  Future studies of its detection efficiency will focus,
for instance, on the longitudinal homogeneity and possible edge
effects.

% -------------------------------------------------------
\section*{CRediT authorship contribution statement}
% -------------------------------------------------------

{\bf M.~Lau{\ss}:} Conceptualization, Formal analysis, Investigation, Methodology, Review \& Editing, Software, Visualization, Writing -- Original Draft.
{\bf P.~Achenbach:} Conceptualization, Formal analysis, Funding acquisition, Investigation, Methodology, Project administration, Review \& Editing, Supervision, Writing -- Original Draft. 
{\bf S.~Aulenbacher:} Review \& Editing.
{\bf M.~Ball:} Review \& Editing. 
{\bf I.~Beltschikow:} Investigation, Review \& Editing, Software.
%{\bf J.~C.~Bernauer:} Review \& Editing.
{\bf M.~Biroth:} Conceptualization, Formal analysis, Investigation, Methodology, Review \& Editing, Visualization. 
{\bf P.~Brand:} Review \& Editing. 
{\bf S.~Caiazza:} Review \& Editing.
{\bf M.~Christmann:} Conceptualization, Investigation, Methodology, Review \& Editing.  
%{\bf E.~Cline:} Review \& Editing. 
{\bf O.~Corell:} Resources, Review \& Editing.
{\bf A.~Denig:} Funding acquisition, Project administration, Review \& Editing.
{\bf L.~Doria:} Funding acquisition, Project administration, Review \& Editing.
{\bf P.~Drexler:} Investigation, Review \& Editing, Software.
%{\bf I.~Fri\v{s}\v{c}i\'c:} Review \& Editing.
{\bf J.~Geimer:} Review \& Editing.
{\bf P.~G\"ulker:} Investigation, Review \& Editing.
%{\bf A.~Khoukaz:} Review \& Editing.
%{\bf M.~Kohl:} Review \& Editing.
{\bf T.~Kolar:} Review \& Editing.
{\bf W.~Lauth:} Conceptualization, Investigation, Resources, Review \& Editing.  
{\bf M.~Littich:} Review \& Editing. 
{\bf M.~Lupberger:} Review \& Editing.
{\bf S.~Lunkenheimer:} Review \& Editing.
{\bf D.~Markus:} Review \& Editing.
{\bf M.~Mauch:} Review \& Editing.
{\bf H.~Merkel:} Funding acquisition, Project administration, Resources, Review \& Editing. 
{\bf M.~Mihovilovi\v{c}:} Review \& Editing.
%{\bf R.~G.~Milner:} Funding acquisition, Review \& Editing. 
{\bf J.~M\"uller:} Review \& Editing.
{\bf B.~S.~Schlimme:} Funding acquisition, Project administration, Review \& Editing. 
{\bf C.~Sfienti:} Funding acquisition, Review \& Editing. 
{\bf S.~\v{S}irca:} Review \& Editing.
%{\bf E.~Stephan:} Review \& Editing.
{\bf S.~Stengel:} Review \& Editing.
{\bf C.~Szyszka:} Review \& Editing.
{\bf S.~Vestrick:} Review \& Editing.
%{\bf Y.~Wang:} Review \& Editing.
%{\bf A. Wilczek:} Review \& Editing.

% -------------------------------------------------------
\section*{Acknowledgments}
% -------------------------------------------------------

The authors would like to thank the MAMI operators, technical staff,
and the accelerator group for their excellent work. We also thank
P.~L.~Cole for language support.

This work was supported by the PRISMA$^+$ Cluster of Excellence
``Precision Physics, Fundamental Interactions and Structure of
Matter'', and by the Helmholtz-Gemeinschaft Deutscher
Forschungszentren (HGF) with a HGF-Exzellenz\-netz\-werk.

% -------------------------------------------------------
%\bibliographystyle{elsarticle-num-names}
%\bibliography{references}

% -------------------------------------------------------

%========================================================
\end{document}